\documentclass[fleqn,10pt]{wlscirep}
\usepackage[utf8]{inputenc}
\usepackage[T1]{fontenc}
\usepackage{mdwlist}  
\usepackage{nameref}
\usepackage{caption}
\usepackage{subcaption}

\title{Robust neuromorphic coupled oscillators for adaptive pacemakers}

\author[1,*]{Renate Krause}
\author[2]{Joanne J.A. van Bavel}
\author[1]{Chenxi Wu}
\author[2]{Marc A. Vos}
\author[3]{Alain Nogaret}
\author[1]{Giacomo Indiveri}
\affil[1]{Institute of Neuroinformatics, University of Zurich and ETH Zurich, Zurich, Switzerland}
\affil[2]{Department of Medical Physiology, Division Heart \& Lungs, University Medical Center Utrecht, Utrecht, The Netherlands}
\affil[3]{Department of Physics, University of Bath Claverton Down, Bath, UK}
\affil[*]{rekrau@ini.uzh.ch}

\keywords{neuromorphic, coupled oscillators, pacemaker, respiratory sinus arrhythmia}

\begin{abstract}
Neural coupled oscillators are a useful building block in numerous models and applications.
They were analyzed extensively in theoretical studies and more recently, in biologically realistic simulations of spiking neural networks.
The advent of mixed-signal analog/digital neuromorphic electronic circuits provides new means for implementing neural coupled oscillators on compact low-power spiking neural network hardware platforms.
However, their implementation on this noisy, low-precision and inhomogeneous computing substrate raises new challenges with regards to stability and controllability.
In this work, we present a robust, spiking neural network model of neural coupled oscillators and validate it with an implementation on a mixed-signal neuromorphic processor.
We demonstrate its robustness showing how to reliably control and modulate the oscillator’s frequency and phase shift, despite the variability of the silicon synapse and neuron properties.
We show how this ultra-low power neural processing system can be used to build an adaptive cardiac pacemaker modulating the heart rate with respect to the respiration phases and compare it with surface ECG and respiratory signal recordings of dogs at rest.
The implementation of our model in neuromorphic electronic hardware shows its robustness on a highly variable substrate and extends the toolbox for applications requiring rhythmic outputs such as pacemakers.
\end{abstract}
\begin{document}
\flushbottom
\maketitle

\thispagestyle{empty}

\section*{Introduction}
\label{sec:introduction}

Computational models of coupled oscillators have been studied extensively in the past~\cite{Ermentrout93,FitzHugh55,Borisyuk_etal95,Terman_Wang95}, and have been used to develop faithful models of Central Pattern Generators (CPGs)~\cite{Kleinfeld_Sompolinsky88,Kopell_Ermentrout88}.
Using both theoretical studies and software simulations, many relevant properties of such systems have been investigated to model mechanisms of locomotion~\cite{Grillner_etal88, Grillner_etal91, Cohen_etal82,  Kopell_Ermentrout88}, respiratory and cardiac rhythms~\cite{Grudzinski_Zebrowski04, Schafer_etal99}, and to drive or control rhythmic movements in robotics~\cite{Gutierrez-Galan_etal20, Lewis_etal00, Lewis_etal03, Tenore_etal04, Vogelstein_etal08}.

In most of these cases, the studies were based on continuous-time coupled differential equations representing simplified neuron models, such as continuous time phase oscillators~\cite{Ermentrout93}, or the FitzHugh-Nagumo~\cite{FitzHugh55} model. 
More recently, these studies were extended to the analysis of coupled populations of physiologically realistic spiking neuron models connected by both excitatory and inhibitory synapses, and were done including the effects of noise and variability~\cite{Chow_Kopell00,Lewis_Rinzel03,Mattia_Del-Giudice04,Brunel_Hakim08,Ponulak_etal06,Zilli_Hasselmo10,Faisal_etal08,Chauhan_etal18}.
These studies revealed how synchronization patterns are strongly influenced by both model and coupling parameters, and provided important insights in the workings of real CPGs. However, they did not take into account all the non-linear and non-ideal effects that could be present in physical implementations, both biological and electronic, such as temperature dependence or limited resolution and dynamic range.

In this work, we propose to go one step further in adding biological realism to these models by implementing coupled oscillators using mixed-signal analog/digital neuromorphic electronic circuits that use the physics of transistors to emulate the biophysics of real neurons and synapses~\cite{Mead90,Chicca_etal14}, and demonstrate a hardware CPG network that operates in real-time and that could therefore be used in closed-loop applications to drive motors and actuators coupled to a wide variety of artificial and biological dynamical systems.

Similar to the original approach of Nagumo, who modeled axonal behaviours using an electric circuit~\cite{Nagumo_etal62}, the approach of using neuromorphic electronic circuits implemented in modern CMOS VLSI technology to build hardware neural circuits aims at directly emulating their dynamics and validating their neural computational properties in real-time compact and embedded systems.
In particular, neuromorphic processors~\cite{Moradi_etal18,Thakur_etal18} are a promising substrate to implement coupled oscillators because they perform parallel processing of large populations of neural circuits in real-time, using ultra low power.
In addition to being a useful research tool for studying the neural circuits they are ideal for implementing closed-loop interactive experiments, with direct and online access to the system parameters that govern their dynamics. This technology is particularly attractive for applications that require compact form-factors, high scalability, low weight, low power and that cannot resort to connecting to remote ``cloud'' computing services for signal processing~\cite{Ma_etal20,Burelo_etal21,Zhao_etal20a,Boi_etal16}.

The main challenges in implementing coupled oscillators with stable and robust behaviour on neuromorphic hardware are the stability and controllability of the model on a noisy substrate.
In mixed-signal analog/digital neuromorphic hardware the neuron and synapse circuits present small deviations from nominal properties (device mismatch) and therefore produce slightly different behaviours in response to the same set of parameters.
Hence, to achieve a stable oscillatory output and a well controlled behaviour as a function of the parameters, the model has to be resilient to inhomogeneities and variability in the synapse and neuron properties.
This is further complicated by the sensitivity of the analog circuits to noise in the input signals and to temperature variations.
This naturally excludes all models which assume that all neuron and synapse equations have identical parameters, and that requires precisely tuned values.

We present a population model for a robust implementation of coupled oscillators on neuromorphic electronic hardware and demonstrate how the frequency and phase shift can be reliably controlled and modulated with tunable parameters.
We validate this approach by applying the proposed neuromorphic system to implement an adaptive cardiac pacemaker proof of concept: we stimulate the hardware oscillator with external input signals describing the respiration phases to modulate the heart stimulation times accordingly and evaluate our results by comparing the generated heart chamber stimulation timings with surface ECG (sECG) and respiratory signal recordings of healthy dogs at rest.

The novel aspects of this work consist in the proposal of a spiking neural network that implements a robust model of coupled oscillators which can work with noisy, low-precision and inhomogeneous computing substrates, and of a real-time CPG system composed of synapse and neuron electronic circuits with biologically plausible dynamics, applied to closed-loop online experiments.

\section*{Results}\label{sec:results}
\subsection*{Spiking neural network model of coupled oscillators on noisy hardware computing substrate}
The mixed-signal analog/digital neuromorphic processor used to validate this model is the Dynamic Neuromorphic Asynchronous Processor (DYNAP-SE) device described in~\cite{Moradi_etal18} (see Fig.~\ref{fig:dynapse_and_architectures}~(a)).
The processor comprises four cores, of 256 neurons each, and 64 dynamic synapses per neuron (see~\cite{Chicca_etal14} for detailed characterization of the analog properties of the synapse and neuron circuits). 
Figure~\ref{fig:dynapse_and_architectures}~(c) illustrates the extend of the device mismatch found on the DYNAP-SE by showing the distribution of synaptic time constant measured on one core over 256 synapses ($30\pm4\,ms$).

\subsubsection*{Neuron population activity}
To maintain high accuracy and robustness despite this device mismatch and the inherent noise present in silicon neurons, we resort to using neuron populations for representing the state variables of the system.
By averaging the spiking activity across multiple neurons, rather than over time, we achieve a high accuracy while avoiding long integration times. 
The activation of such a neuron population is determined by the spiking activity of all neurons in the population combined and calculated as an exponentially moving average activity trace per population ($\tau = 50\,ms$ which roughly corresponds to the time constant of the neurons).
This representation accurately describes the activation of the neuron population as one entity while the temporal smoothing further reduces the impact of the noise found in analog hardware. 
The activity trace itself is calculated off-chip using the spike events generated by the hardware.

A neuron population is considered activated when the population activity trace crosses a predefined threshold value.
This threshold is set to 50\% of the excitatory populations (n=16) and to 25\% of the inhibitory population (n=4) so that it requires more than half of the population to be activated within a small time window to cross the threshold. 
The precise value of these thresholds is not critical: it is sufficient that they are set high enough to avoid false positive activation caused by the spontaneous firing of single neurons rather than the whole population. Furthermore, it is beneficial to choose a rather low threshold value to avoid long delays between the onset of the activation of a population and the detection of it.

\subsubsection*{Neuronal oscillators as the basic building block in network architecture}
\begin{figure}
	\centering
	\begin{subfigure}[b]{0.47\textwidth}
		\includegraphics[width=\linewidth]{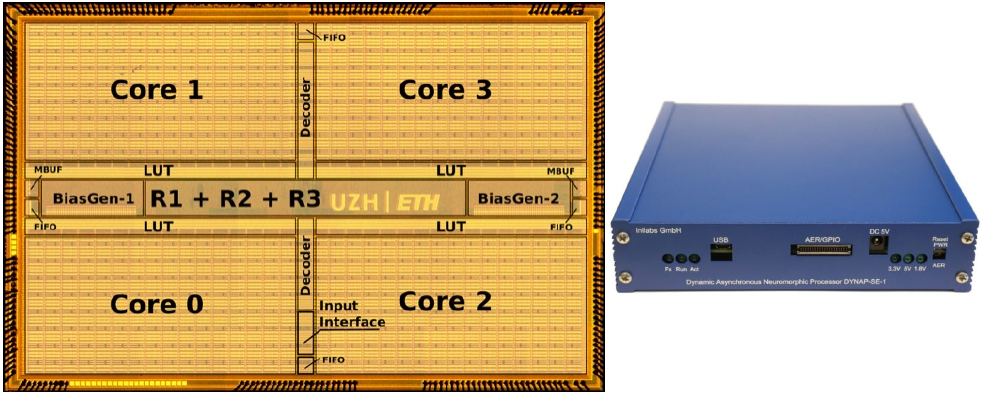}
		\caption{}
	\end{subfigure}
	\begin{subfigure}[b]{0.26\textwidth}
		\centering
		\includegraphics[width=\linewidth]{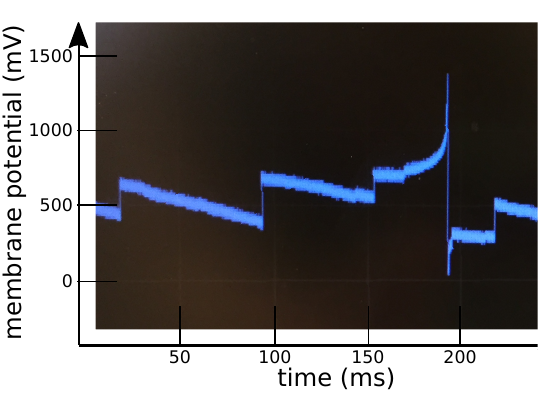}
		\caption{}
	\end{subfigure}
	\begin{subfigure}[b]{0.26\textwidth}
		\centering
		\includegraphics[width=\linewidth]{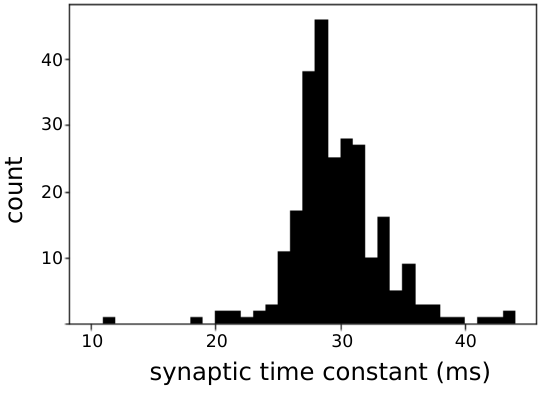}
		\caption{}
	\end{subfigure}
	\begin{subfigure}[b]{0.17\textwidth}
		\centering
		\includegraphics[width=\linewidth]{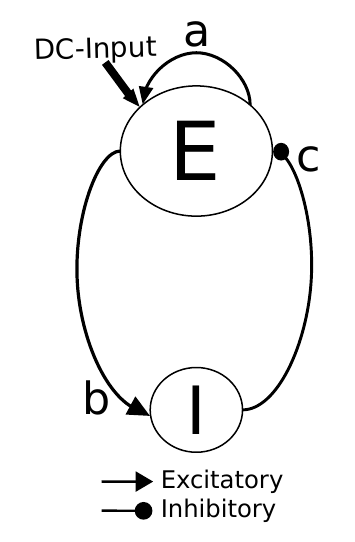}
		\caption{}
	\end{subfigure}
	\begin{subfigure}[b]{0.5\textwidth}
		\centering
		\includegraphics[width=\linewidth]{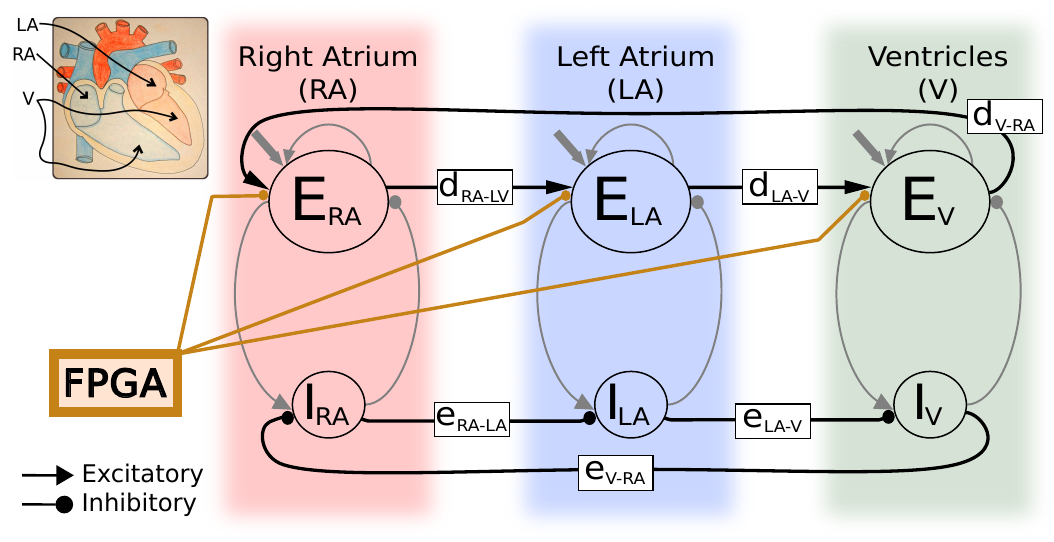}
		\caption{}
	\end{subfigure}
	\begin{subfigure}[b]{0.31\textwidth}
		\centering
		\includegraphics[width=\linewidth]{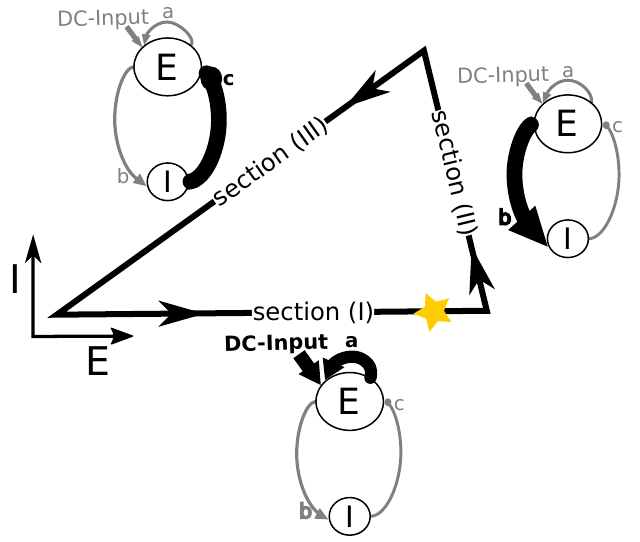}
		\caption{}
	\end{subfigure}
  \caption{\textit{SNN architecture for implementation on neuromorphic processor}
    (a) We implemented our model the mixed-signal spiking neural network processor DYNAP-SE which consists of four cores with 256 neurons each.
    (b) Example membrane trace of one neuron on the DYNAP-SE board. The neuron spikes at around t=200\,ms and receives several incoming spikes which are not sufficient to trigger an event.
    (c) Variability of synaptic time constants on one core of DYNAP-SE board due to device mismatch (n=256).
    (d) A neuronal oscillator is the basic building block in our model.
    It consists of two reciprocally connected inhibitory and excitatory neuron populations.
    (e) In the complete model we use three neural oscillators to model the activation times of the right atrium, the left atrium and the combined ventricles.
    Additionally, we provide external feedback input (respiratory feedback) using the on-board FPGA on the DYNAP-SE board to modulate the heart rate modeling respiratory sinus arrhythmia .
    (f) To tune the model parameters we developed a structured tuning process based on a set of relations between the network function (oscillation properties) and the network parameters and structure. Here we illustrate the relation between the network parameters and the oscillation frequency of a single neuronal oscillator. The connection in bold is used to in- or decrease the speed of a single neural oscillator in the respective section of the phase diagram. (E: activity level of excitatory population; I: activity level of inhibitory population; yellow star: reference point to determine phase shift between oscillators).}
  \label{fig:dynapse_and_architectures}
\end{figure}

We use a neuronal oscillator as a basic building block for our model to provide a modular and flexible network architecture. 
The diagram of a single neuronal oscillator is shown in see Fig.~\ref{fig:dynapse_and_architectures}~(d). 
It consists of an excitatory neuron population E (n=16) and an inhibitory neuron population I (n=4), which are reciprocally connected (connections b and c).
Additionally, the neurons of the excitatory population E receive a constant input current (DC) and are recurrently connected to induce self-excitation (connection a) in~Fig.~\ref{fig:dynapse_and_architectures}~(d).
All connections between the neuron populations are implemented as all-to-all connections.

The reciprocal connections between the excitatory and the inhibitory neuron populations lead to the oscillatory behaviour of the system.
First, the activity of the excitatory population increases due to the constant current input.
Then, the excitatory population activates the inhibitory population which will again inhibit the excitatory population and reset the network.

We combine these single neuronal oscillators to build a system of coupled oscillators consisting of several E/I neuronal populations.
An example system of three coupled oscillators is shown in Fig.~\ref{fig:dynapse_and_architectures}~(e).
We add excitatory connections from every excitatory population to the next oscillator (connections d\textsubscript{RA-LA}, d\textsubscript{LA-V} and d\textsubscript{V-RA})) and inhibitory connections from every inhibitory population to the inhibitory population of the next oscillator (connections e\textsubscript{RA-LA}, e\textsubscript{LA-V} and e\textsubscript{V-RA}) to couple the individual oscillators.
The work of Borisyuk et al.~\cite{Borisyuk_etal95} shows that this architecture leads to antiphase periodic oscillations for any weight of the inhibitory coupling connections and hence further increases the robustness of the model.
Additionally, we add the option to apply an inhibitory input to each excitatory population (shown in orange in ~Fig.~\ref{fig:dynapse_and_architectures}~(e)) to modulate the oscillation frequency depending on an external input signal.
In general, the strength of these connections is subject to the application-specific tuning and hence might either lead to strong or weakly coupled oscillators.
However, due to the limited strength and resolution of the weights present on the chip, the oscillators will be weakly coupled in most cases.

\subsection*{Hardware validation of the model}
The hardware used in this experiment is a custom board that integrates four DYNAPS-SE chips and comprises an internal Field-Programmable Gate Array (FPGA) device that was configured to interface the custom chips to external processing devices and computers.
The board was interfaced to a computer to measure the behaviour of the CPG model, to send control stimuli to the CPG neuron populations, and to change its model parameters.
However, after completing the prototyping phase, used to observe the system output and to change its parameters, the computer can be removed and the chip could run autonomously, in conjunction with the system needed for the closed-loop online application.

\subsection*{Controlling coupled oscillators by relating the network structure to its function}
We studied the relation between the parameters of the SNN model and the network's oscillation properties to develop a structured tuning process to semi-automatically tune our network.
This set of relations provides a link between the network parameters that we can control directly and the oscillation properties that we want to control.
We do not aim to generate a complete mapping between the model parameters and the oscillatory dynamics, but rather focus on the parameters that dominate specific aspects of the network's behaviour that we would like to control.
Specifically, we tune the oscillation frequency, set the phase shift between the coupled oscillators and modulate the frequency of the coupled oscillators with external and variable input signals.

The study of the relation between the parameters of the SNN model and the network's oscillation properties allowed us to tune the network in a semi-automated manner with a computer in the loop rather than tuning the network parameters by hand.
Our tuning process is kept simple and general so that it hopefully can be applied to different SNN models in a diverse set of applications.
In short, first we switch off all not used network components and set all dispensable parameters to default values. 
Then we tune the oscillation frequency of individual oscillators by changing only two specific parameters before we couple the oscillators to tune the system as a whole. 
The complete tuning process is described in detail in the method section.

In this section we describe this set of relations between the network functions and the network structure in the context of three functional aspects of oscillators and explain how these functional properties are controlled by the underlying network structure.

\subsubsection*{Control of oscillation frequency in isolated neuronal oscillators}
The phase diagram in Fig.~\ref{fig:dynapse_and_architectures}~(f) describes the dynamics seen in our neuronal oscillator.
The x- and y-axis show the activity level of the excitatory (E) and inhibitory (I) populations, respectively. 
Over the course of one oscillation, the network follows the trajectory once all around until it gets back to (0/0).
Hence, the oscillation frequency depends on how fast the network is moving along the individual section on this trajectory. 

In this simple setting, we can define which network parameters dominate the network's speed on the different sections along the trajectory.
This relation between the network's function and its structure is illustrated in Fig.~\ref{fig:dynapse_and_architectures}~(f).
The excitatory input to the excitatory population (DC) and the self excitation (Fig.~\ref{fig:dynapse_and_architectures}~(d), connection a) define the speed of how fast the excitatory population gets activated and hence the position of the network along the x-axis (Fig.~\ref{fig:dynapse_and_architectures}~(f), section~(I)). 
A stronger input or larger self excitation accelerates the network along section~(I) and thereby increases its oscillation frequency.
The strength of the connection from the excitatory to the inhibitory population (Fig.~\ref{fig:dynapse_and_architectures}~(d), connection b) sets how fast the inhibitory population gets activated (Fig.~\ref{fig:dynapse_and_architectures}~(f), section~(II)). 
The strength of the connection from the inhibitory to the excitatory population (Fig.~\ref{fig:dynapse_and_architectures}~(d), connection c) defines how fast the excitatory population becomes inhibited once the inhibitory population becomes active again (Fig.~\ref{fig:dynapse_and_architectures}~(f), section~(III)).

Based on this understanding we define which parameters to focus on during the tuning process.
In our case, we use the excitatory input (DC input) as well as the connection from the excitatory to the inhibitory population (Fig.~\ref{fig:dynapse_and_architectures}~(d), connection b) to set the oscillation frequency of isolated oscillators. 
They both have a direct effect on the network speed but in two different sections of the trajectory (see Fig.~\ref{fig:dynapse_and_architectures}~(f)).
We use the excitatory input (DC) to tune the oscillation frequency on a more coarse scale and the connection weight to tune on a more fine scale as the DC input has a stronger effect on the oscillation frequency.
Alternatively, we could adjust the self-excitation of the excitatory population (see Fig.~\ref{fig:dynapse_and_architectures}~(d), connection a).
However, it is beneficial to always set the self excitation to a high level so that the activation spreads quickly within the excitatory population and shows an explosive increase in the population activity level to provide a clear signal of activation.

\subsubsection*{Control of oscillation frequency in coupled neuronal oscillators}
In a system of coupled oscillators the individual oscillators are simultaneously moving along the previously described trajectory while directly pushing or pulling each other along.

The strength of their interaction depends on the strength of the connections between the oscillators and on the activation states of the individual oscillators.
Hence, the interaction strength between two oscillators is not constant over the course of one oscillation period because the effect that one oscillator effectively has on another oscillator also depends on the state of the first oscillator at a given time point (e.g. if an oscillator is not active (at (0/0)), it has no effect on any of the other oscillators).

Hence, the individual neuronal oscillators should all run with the same frequency to achieve a stable system of coupled oscillators.
This ensures that the effect that one oscillator has on another oscillator does not change between two oscillation periods.
Meaning, at the beginning of a new oscillation period all oscillators will be back at the state they were already at one oscillation period ago and hence exhibit the same force on each other over the course of the upcoming oscillation period.
If the oscillators are run with the same frequency, they will also remain in the same order along the trajectory.
This is especially crucial in cases where the oscillators should always be activated in the same sequence (e.g. heart chamber activations in pacemakers).

In general, the synchronization of the individual oscillators is an intrinsic property of nonlinear oscillators whereby the strength of synchronization depends on the strength of synaptic couplings between populations.
However, on the hardware we are not able to set the connection strength high enough to ensure synchronization.
Therefore, it is crucial to set the individual neural oscillations to run at roughly the same frequency already during the tuning process to achieve stable oscillations.

\subsubsection*{Control of phase shift between coupled neuronal oscillators}
We refer to the phase shift as the delays between the activations of the different excitatory populations.
In the phase diagram it is illustrated with the help of a reference point along the trajectory which defines a given activation level of the excitatory population (location on the trajectory in section~(I), illustrated with a yellow star in Fig.~\ref{fig:dynapse_and_architectures}~(f)).
The phase shift is understood as the delays between the time points at which the oscillators pass this reference point.

The phase shift is adjusted by changing the weight of the excitatory connection between two coupled oscillators (see Fig.~\ref{fig:dynapse_and_architectures}~(e), connections d\textsubscript{RA-LA}, d\textsubscript{LA-V} or d\textsubscript{V-RA}).
It is decreased (reduce delay) by increasing the weight from one oscillator to the following oscillator.
So, the following oscillator gets activated faster and thereby reduces their relative delay.
Similarly, the phase shift is increased if the connecting weight is weakened. 

These weight changes have to occur in counteracting pairs to keep the system stable during the tuning of the phase shifts.
In counteracting pairs means that a weight increase in one excitatory connection has to be paired with a weight decrease in another excitatory connection (e.g. increase d\textsubscript{RA-LA}, decrease d\textsubscript{V-RA}). 
This is required because an increase in the excitatory input to an oscillator increases its oscillation frequency.
Hence, these weight changes influence each oscillator's intrinsic frequency and thereby make the system unstable as soon as the difference becomes too large.
By adjusting them in counteracting pairs, we increase an oscillator's speed in one section along the trajectory, while decreasing it in another section and hence hold its intrinsic oscillation frequency steady enough to keep the system stable.

\subsection*{Application to adaptive cardiac pacemakers}
We implement an adaptive cardiac pacemaker model and validate our results with simultaneous recordings of sECG and respiratory signal traces of healthy dogs at rest (data provided from University Medical Center Utrecht, NL).
This allows us to demonstrate the robustness and controllabilty of our approach using neuromorphic electronic circuits in the context of a promising future application.

Our pacemaker model consists of three coupled oscillators generating a three-phase rhythm. 
The activation times of the three oscillators correspond to the activation times of the two atria individually and the two ventricles combined.
We chose to model the two ventricles combined to simplify the model and to take into account that their activations are in general very close in time.
In this model the oscillation frequency corresponds to the heart rate and the phase shift between the individual coupled oscillators describes the delay between the chamber activations.

To implement this network on the DYNAP-SE board we had to set 50 parameters per oscillator, resulting in 150 parameters in total.
Every parameter is shared between all neurons of one core on the DYNAP-SE board whereby there are 25 different parameters per core (each within a range from $0-23\,\mu A$).
In our implementation the parameters are shared for all neurons within one population since we implemented every neuron population on a separate core.
This allowed us to adjust the parameters of every population individually with the semi-automatic tuning process.

In a first step, we show how we achieve stable oscillations of different frequencies and tuned phase shifts.
This is explained by implementing a cardiac pacemaker producing a constant heart rate with tuned delays between the heart chamber activations.
In a second step, we include external signals to modulate the oscillation frequency. 
For this we modulate the heart stimulation times according to the respiration phases in our cardiac pacemaker implementation.
This physiological phenomenon is known as respiratory sinus arrhythmia (RSA) and is characterized by a decrease in the heart rate during exhalation and an increase in the heart rate during inhalation.

\subsubsection*{Model validation}
We validate our hardware model with recordings from healthy dogs at rest, including simultaneous recordings of sECG and respiratory signal traces (data provided from University Medical Center Utrecht, NL).
We focus on recordings of dogs at rest because RSA is found to be more prominent under slow and deep breathing~\cite{ben2012evaluating, OCallaghan_etal20}, and is highly present in resting dogs~\cite{Hamlin_etal66, Mose_etal20}.
We validate our model for RSA with a recording of 2.3\,minutes showing prominent RSA modulation. 
Since the heart rate can be affected by a large set of physiological signals interacting in complicated manners, we aim to present a model which is able to implement the general effect of RSA modulation, and not to reproduce the exact sECG recordings.

\begin{figure}
	\centering
	\begin{subfigure}[b]{0.56\textwidth}
		\includegraphics[width=\linewidth]{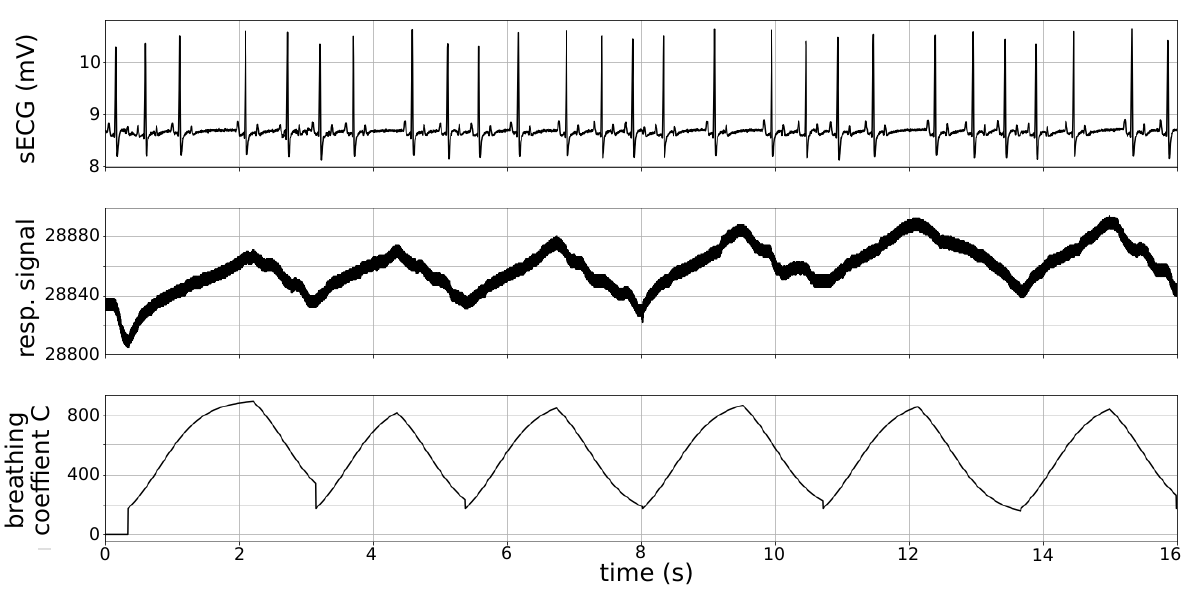}
		\caption{}
	\end{subfigure}
	\begin{subfigure}[b]{0.42\textwidth}
		\includegraphics[width=\linewidth]{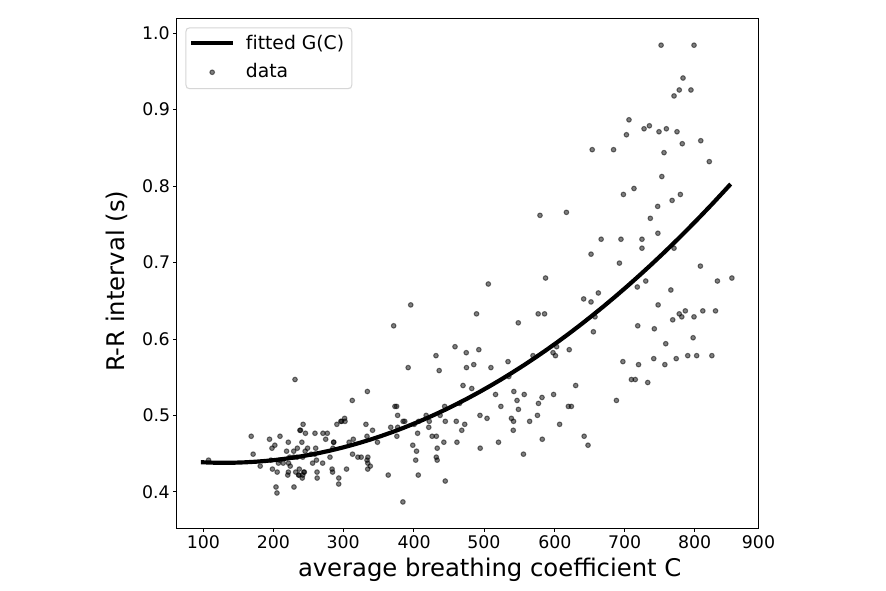}
		\caption{}
	\end{subfigure}
	\caption{\textit{Presence of RSA in sECG and respiratory signal recordings of dogs at rest}
		(a) The first row shows the sECG recordings and the second row the simultaneously recorded respiratory signal traces.
		An increase in the respiratory signal value corresponds to exhalation and a decrease corresponds to inhalation.
		In this recording, RSA is clearly visible in the continuous decrease in the dog's heart rate over the course of the exhalation.
		The third row shows the instantaneous breathing coefficient $C$ during the same period of time.
		The breathing coefficient $C$ is introduced as a new variable to quantify the effect of RSA and can be understood as removal of the linear drift from the respiratory signal trace.
		(b) We quantify the presence of RSA based on the relation between the average breathing coefficient and the R-R interval.
		The plot shows the R-R interval durations plotted against the average breathing coefficient in the corresponding R-R interval. 
		Every point corresponds to one interval extracted from our recordings of a healthy dog at rest.
		The plot shows that a longer R-R interval corresponds to a larger average breathing coefficient.
		The solid line corresponds to a second degree polynomial function G($C$) which is fitted on the retrieved data points.}
	\label{fig:rsa_data}
\end{figure}

To quantify the effect of our RSA modulation, we introduce a variable called the breathing coefficient $C$ which is calculated based on the respiration phases extracted from the respiratory signal traces.
Figure~\ref{fig:rsa_data}~(a) shows an recording sequence of sECG, respiratory signal and the corresponding breathing coefficient to illustrate their relation.
An increase in the respiratory signal value corresponds to exhalation and a decrease corresponds to inhalation.
The plot shows how the heart rate continuously slows down over the course of the exhalation phase.
The instantaneous breathing coefficient $C$ is calculated for every time point and can be understood as removal of the linear drift from the respiratory signal trace.
The breathing coefficient $C$ monotonically increases during exhalation, monotonically decreases during inhalation and is reset to 0 at the beginning of every exhalation phase (see methods for the exact definition of the breathing coefficient $C$).

In our model we aim to reproduce the relation between the R-R intervals (two consecutive heart beats) and the corresponding average breathing coefficient during each R-R interval (average $C$ over corresponding R-R interval).
The average breathing coefficient describes to what extent the current R-R interval happened during a period of inhalation and hence correlates with how strongly the heart rate will be slowed down during this time.
Figure~\ref{fig:rsa_data}~(b) shows the relation between the average breathing coefficients between any two heart beats  plotted against the absolute delay between these two corresponding heart beats for our sECG and respiratory signal recordings of healthy dogs at rest.
In line with the general mode of operation of RSA, it shows that a higher average breathing coefficient is associated with a longer R-R interval (slower heart rate) (fit for G(C): $R^2 = 0.64$ for $G(C)$).

\subsubsection*{Cardiac pacemaker with constant heart rate}
As a baseline mode, we first set up a system of coupled oscillators with a flexible, but constant oscillation period (heart rate) and phase shifts (delays between the heart chamber activations).
Based on all our sECG recordings of healthy dogs at rest, we set our baseline target delay between the activation of the right and left atria to 15\,ms, between the left atrium and the ventricles to 110\,ms, and to 430\,ms for the delay from the ventricles until the start of a new cycle by the activation of the right atrium (overall oscillation period of 555\,ms).

\begin{figure}
	\centering
	\begin{subfigure}[b]{0.53\textwidth}
		\includegraphics[width=\linewidth]{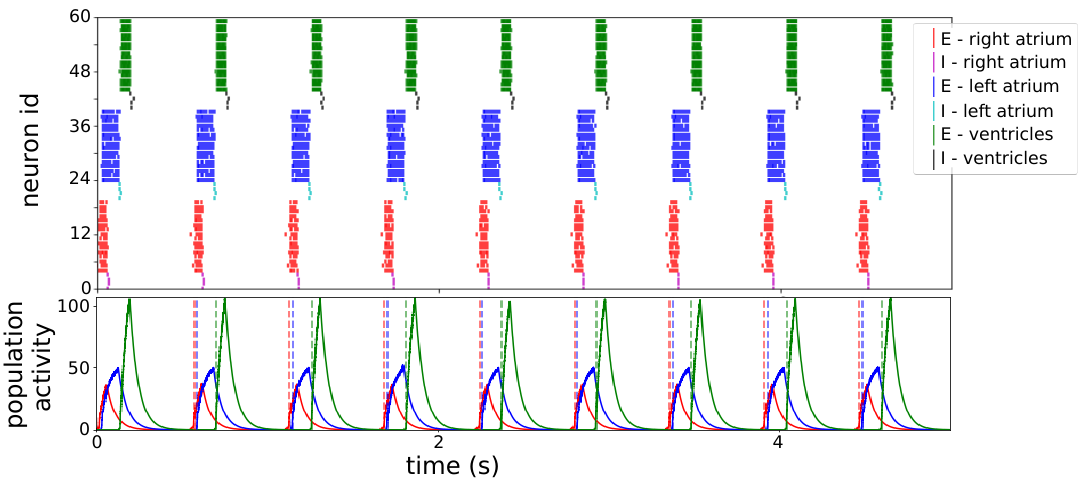}
		\caption{}
	\end{subfigure}
	\begin{subfigure}[b]{0.23\textwidth}
		\centering
		\includegraphics[width=\linewidth]{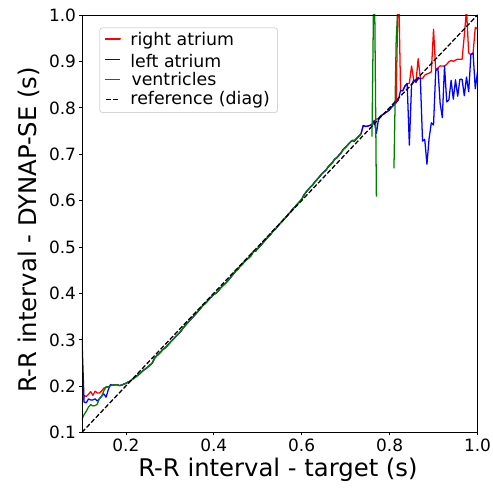}
		\caption{}
	\end{subfigure}
	\begin{subfigure}[b]{0.23\textwidth}
		\centering
		\includegraphics[width=\linewidth]{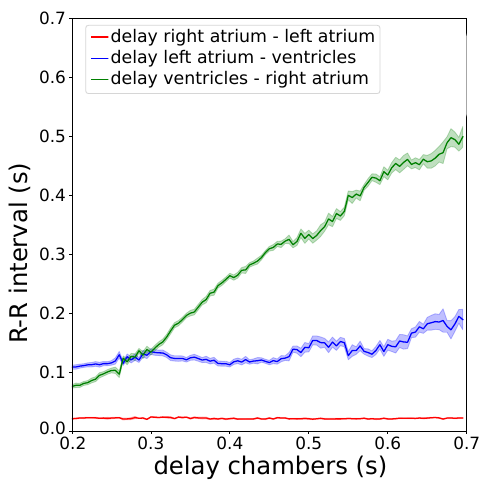}
		\caption{}
	\end{subfigure}
	\caption{\textit{Cardiac pacemaker implementation with constant heart rate}
		(a) The raster plot shows the events of all neurons of the three coupled neuronal oscillators on the DYNAP-SE board over 5\,seconds.
		The inhibitory and excitatory populations of each oscillator are presented in different colors.
		(b) The events are converted into activity traces for each population. Here we only show them for the three excitatory populations. The stimulation time of each heart chamber is defined as the time point when the activity traces of the respective excitatory population crosses a predetermined and fixed threshold value.
		(c) After the initial iterative tuning process, we are able to set the systems oscillation frequency explicitly by fitting a mapping function. The x-axis shows the desired oscillation frequency, the y-axis shows the oscillation frequency measured on the DYNAP-SE board after setting the biases based on the fitted mapping function. This plot shows that we are able to explicitly set the oscillation period of the coupled system to any value between 200-700\,ms.
		(d) The delay (phase shift) between the activation of the heart chambers smoothly adjusts itself to the explicitly set heart rate without any further tuning.}
	\label{fig:cardiac_CPG}
\end{figure}

We implemented the network architecture of three coupled oscillators on the DYNAP-SE board, performed the iterative tuning process we described, and achieved these target delays.
The average delays between the activations of the right and left atrium was $17\pm0.3\,ms$, between the left atrium and the ventricles was $108\pm3\,ms$ and between the ventricle and the right atrium was $431\pm3\,ms$.
The average oscillation period (measured over 30\,s) is $556\,ms$.
Figure~\ref{fig:cardiac_CPG}~(a) shows the spikes of all neurons in the six neuron populations of the three-phase cardiac pacemaker model measured on the DYNAP-SE board.
It shows the robust oscillatory activity and interaction of all three coupled neuronal oscillators. 
Below in Fig.~\ref{fig:cardiac_CPG}~(b) we plot the resulting decaying activity traces of the three excitatory neuron populations.
These activity traces are used to extract the heart chamber activation timings which are indicated with dashed black lines.

After the initial, iterative tuning process the system is in a stable regime and we can set the heart rate (oscillation frequency) in an explicit manner.
This is required to change the heart rate without additional tuning and only a minor delay.
The explicit mapping is based on the knowledge from the iterative tuning process where the oscillation frequency of an isolated neuronal oscillator is tuned by adjusting the excitatory input to its excitatory population.
It is obtained by collecting a small dataset on the hardware and then fitting a function from the input strength to the resulting oscillation period for each oscillator in isolation (see methods for more details).
These mapping functions then allow us to adjust the heart rate by setting each neuronal oscillator to the common target frequency.

The mapping functions enable us to directly set the oscillation period to any value between 200 - 700\,ms (see Fig.~\ref{fig:cardiac_CPG}~(c)).
Outside of this range the oscillation becomes unstable. 
However, this covers the oscillation periods required to model heart chamber activations and hence is sufficient.

Furthermore, even though we do not tune the delay between the heart chambers (phase shift) when setting the heart rate explicitly, we observe that the delays change continuously and stay in a physiologically plausible regime.
Figure~\ref{fig:cardiac_CPG}~(d) shows the delays between the heart chambers for different heart rates.

\subsubsection*{Adaptive cardiac pacemaker modeling respiratory sinus arrhythmia}
Next, we added the respiratory feedback signals to modulate the heart rate to achieve the previously extracted relation between the breathing coefficient $C$ and the R-R interval $T_h$ (see Fig.\ref{fig:rsa_data}~(b)).
The respiratory feedback is provided to the excitatory populations as inhibitory, spiking input (see Fig.~\ref{fig:dynapse_and_architectures}~(e) for complete network architecture).
The spike rate of the inhibitory input is tuned in a semi-automatic manner with a computer in the loop (see methods for more details).
In the tuning procedure we fit a function $H(C)$ which defines the input spike rate $r_{inh}$ for every time point based on the instantaneous breathing coefficient $C$.

\begin{figure}
    \centering
	\begin{subfigure}[b]{0.56 \textwidth}
		\includegraphics[width=\linewidth]{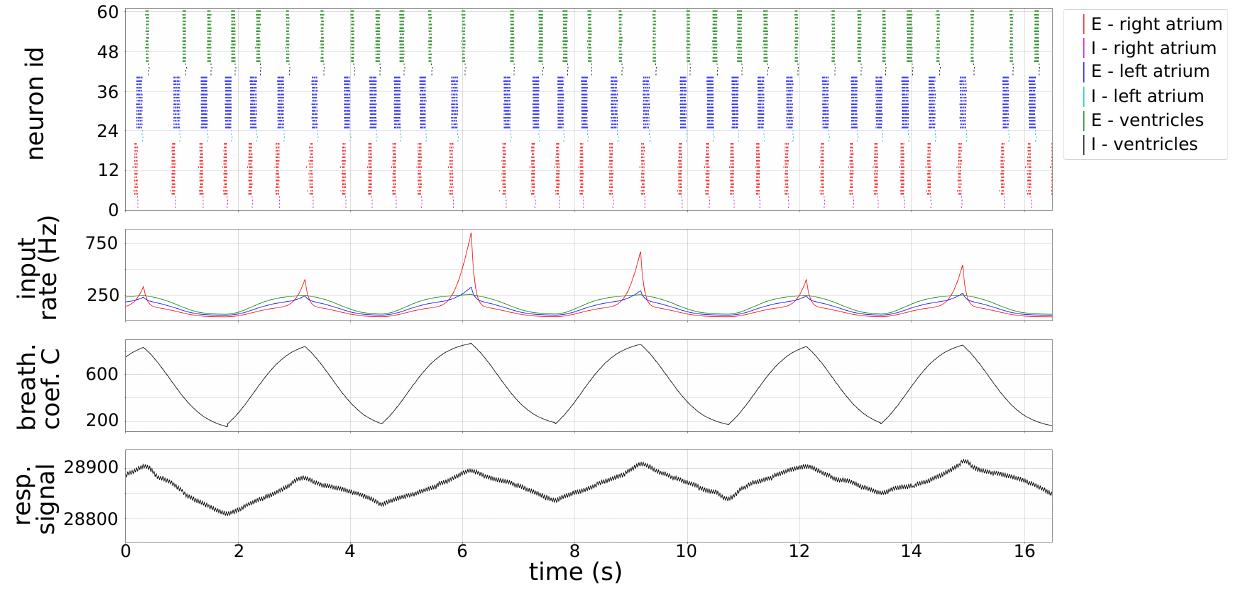}
		\caption{}
	\end{subfigure}
	\begin{subfigure}[b]{0.4\textwidth}
		\includegraphics[width=\linewidth]{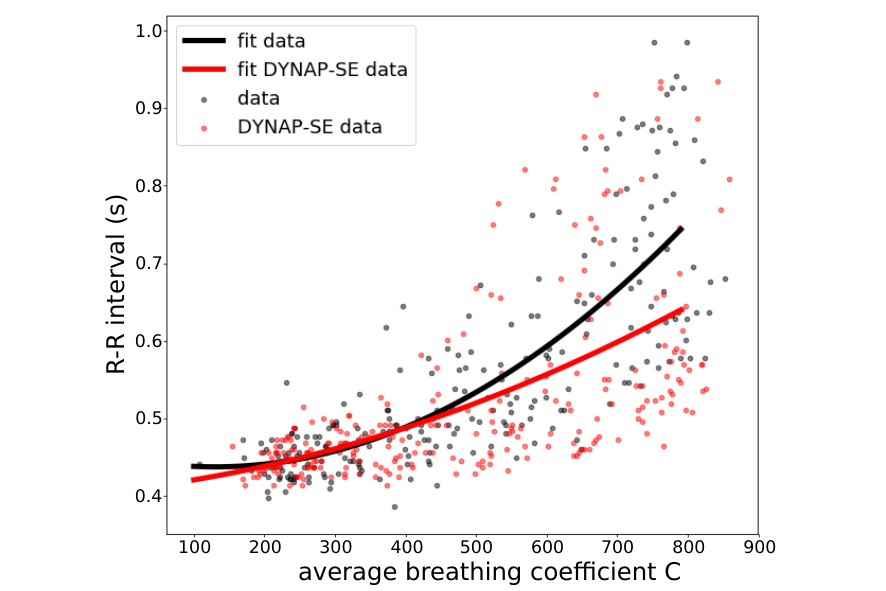}
		\caption{}
	\end{subfigure}
    \caption{\textit{Adaptive cardiac pacemaker implementation with respiratory feedback}
        (a) We tested our implementation of an adaptive cardiac pacemaker on the DYNAP-SE board by providing real respiratory feedback signals.
        The last row shows an example trace of the raw respiratory signal recording over 16\,seconds.
        The rows above show the corresponding breathing coefficient trace $C(t)$ and the resulting input spike frequencies which define the strength of the inhibitory input modelling the respiratory feedback.
        The top row shows the system's behaviour while receiving the respiratory feedback.
        Qualitatively, this shows a continuous decrease of the heart rate during exhalation and an increase during inhalation modeling the effect of RSA.
        (b) To evaluate our model more quantitatively, we extract the relation between every R-R interval and the breathing coefficient averaged over the period between the two respective R-peaks.
        The plot shows that our model reproduces the same relation that is also observed in physiological recordings.}
        \label{fig:adaptive_cardiac_CPG}
\end{figure}

We tested our model by providing it with realistic, respiratory feedback signals based on the respiratory signal recordings of a dog at rest. 
Figure~\ref{fig:adaptive_cardiac_CPG}~(a) shows the respiratory signal values, the derived breathing coefficients and the spike frequency of the provided respiratory feedback over time.
The resulting spiking activity of the coupled oscillators is shown in the top row and shows that the heart rate continuously de- and increases based on the respiration phases seen in the respiratory signal in the last row. 
This shows qualitatively that our model reproduces heart rate modulations which are similar to what is observed for RSA.

We checked this qualitative result by testing whether our model is also able to reproduce the relation between the breathing coefficient $C$ and the R-R interval.
To fit $G(C)$ we provided the network with respiratory feedback for 140\,s and extracted the delays between all generated heart beats and the average breathing coefficients for each interval respectively.
We fitted a second degree polynomial function $G_{Dyn}(C)$ on the data collected from the DYNAP-SE board and achieved a fit $R^2=0.78$ on $G(C)$.
Figure~\ref{fig:adaptive_cardiac_CPG}~(b) shows the data collected on the DYNAP-SE board as well as the resulting fits compared to the physiological data in comparison.
We are able to match the short delays well, but for higher average breathing coefficient, the generated heart rate tends to be a bit too fast.

\section*{Discussion}\label{sec:discussion}
In this work we present a robust implementation of neural coupled oscillators on neuromorphic electronic hardware and show how it can be tuned to implement different types of CPG models.
The modular architecture of the model and the general nature of the parameter tuning procedure allow the implementation of a wide range of dynamics.
This flexibility makes this model a useful building block for a large set of applications that require stable and well controllable oscillatory behaviour (e.g. for driving locomotion in robotics.

We demonstrate how our model can be applied to implement an adaptive cardiac pacemaker that modulates the heart rate based on physiological breathing patterns.
Our model consists of three coupled neural oscillators and the spikes produced by the hardware neuromorphic pacemaker model can be used to determine the heart chamber stimulation times of the right atrium, the left atrium and the two ventricles combined. 
However, the model architecture is set up to be general so that any combination of heart chambers can be modeled.
Additionally, our model receives respiratory feedback which allows it to adapt the heart rate to the respiration phases.
We evaluated our results with sECG and respiratory signal recordings of dogs at rest.

We showed how these circuits can be tuned to exhibit realistic dynamics with physiological time constants.
The results presented demonstrate a proof of concept based on a general purpose neuromorphic processor (the DYNAP-SE chip).
Networks of similar size implemented on the same DYNAP-SE chip have been estimated to dissipate very little power, well below 1\,mw~\cite{Bauer_etal19}. 
Therefore, by integrating the network presented in this work, using the same silicon synapse and neuron circuits, in a dedicated full-custom device optimized for generating rhythms of the types described here, it will be possible to build ultra-low power adaptive pacemakers that can be modulated by physiological feedback signals, such as the breathing patterns (e.g., see the EU H2020 CResPace project~({\url{https://crespace.eu/}}).

\section*{Methods}
\label{sec:methods}
\subsection*{Parameter tuning procedure for constant oscillation frequency and phase shift}
Here we describe the tuning process to set a system of coupled oscillators to run with a given constant oscillation frequency and phase shift.
\subsubsection*{Iterative parameter tuning}
The iterative, and more precise, tuning process consists of four steps and was performed semi-automatically with a computer in the loop.
\begin{enumerate}[noitemsep,topsep=0pt]
	\item \textit{Switch off not used components}\\
	The DYNAP-SE board implements spike-frequency adaptation as well as four types of synapses.
	We do not use adaptation and nor all types of synapses in our model and therefore set their parameters so that they do not interfere.
	
	\item \textit{Set dispensable parameters to default values}\\
	Every neuron and synapse comes with three parameters to tune: a time constant, a gain and a weight term (see \cite{Chicca_etal14} for the corresponding circuit diagram).
	All three parameters act on the amount of current that the neurons receive as input.
	We focus on only one of them (weight parameter) to simplify the tuning process and set the others (time constant and gain) to a medium, default value.
	
	\item \textit{Tune frequency of individual oscillators}\\
	First, we tune the frequency of all neural oscillators separately.
	This allows us to take the device mismatch on the DYNAP-SE board into account and partially correct for it.\\
	First, we set the DC current which is provided as a constant current input to the excitatory population.
	It serves as a coarse tuning step and is adjusted iteratively until the oscillation period measured is not more than 50\,ms off the target oscillation period (increase DC if oscillation frequency too low, decrease DC if oscillation frequency too high).
	Second, we fine tune the oscillation frequency by iteratively adjusting the weight of the connection from the excitatory to the inhibitory neuron population (increase weight if oscillation frequency too low, decrease weight if oscillation frequency too high)).
	
	\item \textit{Tune phase-shift between oscillators}\\
	The delays between oscillators is tuned by iteratively adjusting the weight connecting the excitatory populations (see Fig.~\ref{fig:dynapse_and_architectures}~(e), connections d\textsubscript{RA-LA},d\textsubscript{LA-V} and d\textsubscript{LA-V}).
	Thereby it is important to preserve the initial oscillation frequency of the coupled oscillators to keep the system in a stable regime.
	This is achieved by always changing the weight parameters in counteracting pairs (increase one coupling connection weight, decrease another).
\end{enumerate}

\subsubsection*{Explicit parameter tuning}
This method step seeks to calibrate the frequency to stimulation current response of individual oscillators to change the oscillation frequency explicitly rather than iteratively.
For this, we bring the system into a stable regime by following the iterative tuning process described above and then probe the DYNAP-SE board with DC input currents from 0 to 200\,nA to extract the resulting oscillation frequencies.
Based on this collected dataset, we fit a double exponential decay function
\begin{equation}
f(T) = x_1 * \exp^{-x_2 * T} + x_3 * \exp^{-x_4*T}
\end{equation}
which provides a direct mapping from the desired oscillation period $T$ to the corresponding DC input current $f(T)$.
Note that the tuning accuracy will be lower when using the explicit tuning compared to the iterative tuning.

\subsection*{Tuning of heart rate modulation in adaptive cardiac pacemaker}
\subsubsection*{Extraction of relation between heart rate and breathing rhythm}
To describe the relation between the sECG and respiratory signal recordings quantitatively we introduce a variable called breathing coefficient $C$ which describes the weight with which the heart rate is slowed down at any time point.
The breathing coefficient $C$ corresponds to the external signal $C$ which modulates heart rate over time.
Its value is always reset at the beginning of the exhalation phase.
During the exhalation phase itself the breathing coefficient increases monotonically, representing the increasing slow-down in the heart rate over the course of the exhalation.
During inhalation it decreases monotonically to model the decrease in the slow-down of the heart rate.
In general, a higher average breathing coefficient corresponds to a slower heart rate.

Furthermore, the breathing coefficient is robust to noise or large drifts in the respiratory signal recording since it only depends on the time of the exhalation and inhalation phase onset (peak in respiration signals) but not the absolute value.

More formally, we describe this relationship as $G(C)$, where we extract the average breathing coefficient $C$ for every interval between any two R-peaks and relate it to the absolute delay between these two R-peaks $T_h$: 
\begin{equation} \label{eq:G(C)}
T_h = G(C)
\end{equation}
\begin{equation}
G(C) = x_1*C^2 + x_2*C + x_3
\end{equation}
$G(C)$ is modeled as a second degree polynomial with parameters $x_1$, $x_2$ and $x_3$.
We optimized the parameters of the logistic function of describing the breathing coefficient $C$ to achieve an optimal fit for a second degree polynomial function $G(C)$.
In our hardware model, we aim to reproduce $G(C)$ (Eq.~\ref{eq:G(C)}).
For this, the strength of the inhibitory, respiratory feedback input has to be tuned accordingly.

\subsubsection*{Parameter tuning of external, inhibitory input to modulate the oscillation frequency}
The strength of the inhibitory, respiratory feedback input to the excitatory populations depends on the DYNAP-SE parameter settings for the inhibitory synapse (synaptic weight, synaptic time constant and synaptic gain term) as well as on the firing rate of the inhibitory input.
To simplify the tuning process we focus on adjusting the inhibitory input spike rate and set the synapse parameters to a medium, default value.

Our goal is to fit a function $H$ which defines the inhibitory input spike rate $r_{inh}$ based on the external signal $C$ to induce delays which match the desired oscillation periods $T_h$ described by a function $G(C)$ (Eq.~\ref{eq:G(C)}).
\begin{equation} \label{eq:H(T_h)}
r_{inh} = H(G(C))
\end{equation}
with Eq.~\ref{eq:G(C)}:
\begin{equation} \label{eq:H(T_h2)}
r_{inh} = H(T_h)
\end{equation}

To obtain the function $H$, we perform the following four tuning steps in a semi-automatically manner with a computer in the loop.
\begin{enumerate}[noitemsep,topsep=0pt]
	\item \textit{Set oscillation frequency of every neuronal oscillator to maximal oscillation frequency needed}\\
	This is required because the inhibitory input can only decrease the oscillation frequency but not increase it.

	\item \textit{Obtain approximation for $H$ for constant input signal}\\
	In this tuning step we fit a rough approximation of $H$ which is further fine-tuned in step 3. 
	To obtain this approximation, we probe the DYNAP-SE board by sending inhibitory input of constant spike rates and extract the resulting delays in every oscillator's oscillation period.
	Based on this dataset, we fit $r_{inh} = H(T_h)$ for every neuronal oscillator separately.
	
	\item \textit{Fine-tune $H$ with time-varying input spike rates}\\
	To improve the approximation of $H$, we create and run every oscillator separately with time-varying input spike trains based on the time-varying external signal $C$.
	The approximation of $H$ yielded good results for small external signal $C$, but created too weak inhibitory input firing rates for external signal $C$ so that no long delays were achieved.
	To correct for this, we check for every oscillator what maximal delay $T_{thr}$ it is able to achieve when stimulated with the realistic input spike trains.
	Then, we add an exponentially increasing term to all inhibitory inputs with an input signal $C$ corresponding to longer delays than $T_{thr}$.
	The resulting function to set the firing rate of the inhibitory input is:\\
	\begin{equation}
	r_{inh} = H(T_h) + exp((T_h-T_{thr})k)
	\end{equation}
	The factor k is determined by increasing it iteratively until the longest delays seen in the data (900 - 1000\,ms) are achieved with the realistic, time-varying input spike rates.
	
	\item \textit{Fine-tune $H$ with time-varying input spike rates in coupled oscillators}\\
	The last tuning step serves to harmonize the $H$ functions of the individual oscillators and to take the additional, excitatory coupling input per oscillator into account.\\
	If an oscillator receives only minor input from the other coupled inputs, we can assume that the previous tuning steps were sufficient.
	In this case, we will not adjust its parameters any further and use this oscillator as a reference point for the other oscillators.
	In our system, this is the oscillator representing the right atrium since it receives only minor coupling input because of the long delay between the activation of the ventricle to right atrium.\\
	To tune the other oscillators, we introduce another constant scaling factor $s$ which acts on the external signal $C$ for each oscillator individually and allows us to in- or decrease the produced firing rates to speed-up or slow-down a neuronal oscillator.
	\begin{equation}
	r_{inh} = (H(T_h) + exp((T_h-T_{thr})k))*s
	\end{equation}
	$s$ is again adjusted iteratively, by running the system of coupled oscillators with the realistic, time-varying input spike trains and checking that the neuron populations are activated in the correct order.
	We compare the pairwise activation times of the oscillators. If the order is not preserved, we increase the scaling factor of the oscillator that spiked too early and decrease it for the oscillator that spiked too late.\\
	We perform the optimization of the scaling factor sequentially for the different oscillator pairs.
	Meaning, first we compare the timing of the leading oscillator to the directly preceding one and only if this comparison is fine we move on to the following oscillator pair.
\end{enumerate}

To run the final model we create the realistic breathing coefficient traces $C(t)$ based on the respiratory signal recording. 
Then, we convert these traces $C(t)$ into input spike trains based on every oscillator's function $H((G(C(T)))$ and run the coupled system with these inhibitory inputs.

\subsection*{Recordings of sECG and respiratory signals from dogs}\label{subsec:rec_sECG_LI}
\subsubsection*{Animal Handling}\label{subsubsec:animal_handling}
Handling of the dogs was in accordance with the Directive 2010/63/EU of the European Parliament, the Council of 22 September 2010 on the protection of animals used for scientific purposes, and the Dutch law on animal experimentation.
Experiments were approved by the Committee for Experiments on Animals of Utrecht University.\\
Dogs were housed in conventional kennels of $\pm\,8\,m\textsuperscript{2}$ with wooden bedding.
During the day, the animals went to an outdoor pen for a couple of hours.
The dogs had ad libitum access to water, received food pellets twice a day, and their well-being was checked daily. 

\subsubsection*{Data recording}\label{subsubsec:data_recording}
Healthy and awake adult purpose-bred Mongrel dogs (Marshall, United States) got acquainted to stand in a custom-build closed chamber of plexiglass with a normal air flow (200 L/min).
The Hexoskin shirt (Carré Technologies Inc, Montreal, Canada) was placed around the thorax of the dog and the sECG and respiratory signals, that were recorded by the electrodes, were checked via the Hexoskin application on a mobile phone.
The sECG and respiratory signals were recorded for 5 minutes.

\subsubsection*{Extraction of heart chamber activation timings}
Five sECG and respiratory signals from healthy dogs (n=3) were included in this study (detailed information can be find in section~\ref{subsubsec:animal_handling} and \ref{subsubsec:data_recording}).
The sECG recordings allowed us to extract the average delays between the activations of atria and ventricles. 
The delay between the left and right atrium could not be extracted from the recordings but is estimated to be around 15\,ms~\cite{Feld_Shahandeh-Rad92, Scherlag_etal72, Sweeney14}.

\subsection*{Code availability}
We used Python and the Python API \textit{Samna legacy} (\url{https://pypi.org/project/samna/}) to configure the DYNAP-SE board, tune the network parameters and perform the data analysis.
Extraction of R-peaks (ventricular activation on sECG) was done with BioSPPy~\cite{Carreiras_etal15}.
The code to tune and run the system of coupled oscillators on the DYNAP-SE board is available at: \url{https://gitlab.com/rekrau/neuromorphicOscillatorsForPacemakers}.

\bibliography{biblioncs, bibliork}

\section*{Acknowledgements}
The authors would like to thank SynSense AG for their support on the DYNAP-SE software, 
Matthew Cook for helpful discussions, 
H.D.M. Beekman (Department of Medical Physiology, Division Heart \& Lungs, University Medical Center Utrecht, Utrecht, The Netherlands) and A. Schot (Department of Clinical Sciences, Division of Anatomy and Physiology, Faculty of Veterinary Medicine, Utrecht University, The Netherlands) for their support in all animal experiments and T. Takken (Department of Exercise Physiology, Child Development \& Exercise Center, Wilhelmina Children’s Hospital, University Medical Center Utrecht, Utrecht, The Netherlands) for providing the Hexoskin shirt.
This work was supported by the European Union's Horizon 2020 FET project CResPACE (No. 732170).

\section*{Author contributions statement}
R.K. and C.W. designed the model.
R.K. implemented the model in hardware and performed the analyses.
G.I. co-designed the hardware platform and supervised R.K. and C.W..
J.v.B. performed the animal experiments.
M.V. supervised J.v.B..
A.N. initiated the concept of synthetic CPGs as adaptive pacemakers and secured project funding.
All authors contributed and reviewed the manuscript.

\section*{Additional information}
The authors declare no competing interests.

\newpage

\section*{Supplementary Information} \label{supplementary_information}
\subsection*{Neuron and synapse model on neuromorphic processor}
The equations that describe the behaviour of the neuron and synapse circuits have been derived from the analog circuits~\cite{Chicca_etal14}. In particular, the equation that describes the subthreshold current $I_{mem}$ representing the membrane potential of the neurons on the DYNAP-SE board is, to first order approximation:
\begin{align}
  & \tau \frac{d}{dt} I_{mem} + I_{mem} = I_{mem_\infty} - I_{ahp} +f(I_{mem}) \label{eq:imem} \\
  & \tau_{ahp}\frac{d}{dt}I_{ahp} + I_{ahp}  = I_{ahp_{\infty}} \delta(t)  
\end{align}
where $I_{ahp}$ is the after-hyper polarizing current that implements the neuron's spike-frequency adaption mechanism, $\tau$ and $\tau_{ahp}$  are the neuron's and the adaptation time constants respectively, $\delta(t)$ is the impulse function that is 1 when the neuron spikes and 0 otherwise.
The term $f(I_{mem})$ is an exponential function with positive exponent (see~\cite{Chicca_etal14} for details). As a consequence, the silicon neuron circuits implements an ``adaptive exponential integrate and fire'' (AdExp) neuron model~\cite{Brette_Gerstner05}.

The other parameters in the equation above are:
\begin{align}
  \tau & \triangleq \frac{C_{mem}U_T}{\kappa I_\tau} & \tau_{ahp} & \triangleq \frac{C_pU_T}{\kappa I_{\tau_{ahp}}} \\
  I_{mem_\infty} & \triangleq \frac{I_{g}}{I_\tau}(I_{in} - I_{ahp} - I_\tau) &   I_{ahp_\infty} & \triangleq \frac{I_{g_{ahp}}}{I_{\tau_{ahp}}}I_{Ca} \label{eq:imeminf}
\end{align}

where

$C_{mem}$ and $C_p$ are the capacitors used in the circuit to store the membrane potential variable and the spike-frequency adaptation one. The term  $U_T$ represents the thermal voltage and $\kappa$ the subthreshold slope factor. The currents $I_{\tau}$ and $I_{\tau_{ahp}}$ are hyper-parameters that can be used to set the corresponding time constants, and $I_{g}$ and $I_{g_{ahp}}$ are extra free parameters that can be sued to set global gain terms.

Similarly, to first order approximation, the synaptic input currents $I_{syn}$ that are being summed and sent in input to the neuron (as $I_{in}$ in Eq.~\ref{eq:imeminf}) can be described as:
\begin{equation}
	\tau\frac{d}{dt}I_{syn} + I_{syn} = \frac{I_g I_w}{I_\tau}
\end{equation}
where $I_w$ is the synaptic weight current, $I_g$ a global synaptic scaling term common to all afferent synapses to the same neuron, and $I_\tau$ the current used to set the synapse time constant $\tau$, as defined for the neuron.

\end{document}